\documentstyle[prd,aps,epsfig]{revtex}
\begin{document}
%
%
\def\ov{\over}
\def\l{\left}
\def\r{\right}
\def\be{\begin{equation}}
\def\ee{\end{equation}}
\draft
\title{On gravitational waves emitted by an ensemble\\ 
of rotating neutron stars}
\author{Adalberto~Giazotto}
\address{Istituto Nazionale di Fisica Nucleare, Sezione di Pisa \\
Via Livornese, 1291, I-56010 S.~Piero a Grado (Pisa), Italy}
\author{Silvano Bonazzola and Eric Gourgoulhon}
\address{D\'epartement d'Astrophysique Relativiste et de Cosmologie \\
  UPR 176 du C.N.R.S., Observatoire de Paris, \\
  F-92195 Meudon Cedex, France}
\date{7 October 1996; accepted for publication in {\sl Phys. Rev. D};
preprint astro-ph/9611188}
\maketitle

\begin{abstract}
We study the possibility to detect the gravitational wave
background generated by all the neutron stars in the Galaxy 
with {\em only one} gravitational wave interferometric detector.
The proposed strategy consists in squaring the detector's output and
searching for a sidereal modulation. The shape of the squared signal is 
computed for a disk and a halo distribution of neutron stars. The required
noise stability of the interferometric detector is discussed.
We argue that a possible population of old neutron stars, originating
from a high stellar formation rate at the birth of the Galaxy and not
emitting as radio pulsars, could be detected by the proposed technique
in the low frequency range of interferometric experiments. 
\end{abstract}


\section{Introduction}

Rotating neutron stars (NS) are possible astrophysical sources of 
gravitational radiation in the frequency range of the interferometric
detectors LIGO, VIRGO and GEO600 currently under construction 
\cite{BonaM94,MarcL96,CiufF96}. Indeed, provided it deviates from axisymmetry,
a rotating NS emits continuous wave (CW) gravitational radiation mainly at
its rotation frequency and twice this frequency \cite{Ipser71}. 
The non-axisymmetric shape of a NS can be caused by 
anisotropic stresses from the nuclear interactions,
irregularities in the solid crust (``mountains'') \cite{AlpaP85,Haens96}, 
the internal magnetic field \cite{GalTT84,BonaG96}, 
some precessional motion \cite{ZimmS79},
or the development of triaxial instabilities (for the most rapidly rotating NS):
the Chandrasekhar-Friedman-Schutz instability \cite{Wagon84,LaiSh95}
and the MacLaurin-Jacobi type instability induced by viscosity
\cite{IpseM84,BonFG96}.  
Estimates of upper bounds on the individual 
gravitational wave amplitude from a sample of 334 observed NS (radio
pulsars) have been provided by Barone et al. \cite{BaMPR88} (see
ref.~\cite{VBCDG96} (resp. ref.~\cite{GourB96}) 
for recent values based on a sample of 558 (resp. 706) pulsars). 
The amplitude of gravitational waves (GW) emitted by 
a rotating NS can be expressed in terms
of the NS rotation period $P$, its distance to the Earth $r$, its moment
of inertia $I$ about the rotation axis and its ellipticity (triaxial 
deformation) $\epsilon$ as (cf. Eq.~(\ref{e:Ai}) below)
\be \label{e:h0,num}
    h_0 = 4.21\times 10^{-24} \ \l( {{\rm ms}\ov P} \r) ^2
	\l( {{\rm kpc}\ov r} \r) 
	\l( {I\ov 10^{38} {\ \rm kg\, m}^2} \r)
	\l( {\epsilon \ov 10^{-6} } \r) .
\ee
The crucial parameter entering this formula is the ellipticity $\epsilon$.
Its value depends on the physical mechanism that makes the star 
non-axisymmetric (cf. the references given above) and is highly uncertain. 
Upper bounds on $\epsilon$ can however be derived from the observed
slowing down ($\dot P$) of pulsars by assuming that this latter 
is entirely due to
the loss of angular momentum by gravitational radiation. 
Let us note that this provides an absolute upper bound;  
most of the $\dot P$ is usually thought to result instead from losses via
electromagnetic radiation and/or magnetospheric 
acceleration of charged particles --- at least for Crab-like pulsars.
The maximum values of $\epsilon$ obtained in this way, as well as the
corresponding maximum values of $h_0$, are given in Table~\ref{t:hmax}
for five pulsars. 
The first three of them correspond to the three highest values of 
$h_{0,\rm max}$
among the 706 pulsars of the catalog by Taylor 
et al.\cite{TayML93,TaMLC95}. The two remaining entries 
are two millisecond pulsars: the second fastest one, PSR B1957+20,
and the nearby pulsar PSR J0437-4715. 
The figures $h_{0,\rm max} \sim 10^{-24}$ for the Crab-like pulsars 
(three first entries
in Table~\ref{t:hmax}) are almost certainly too optimistic
because, as already said, electromagnetic phenomena can be invoked 
to explain most of, if not all, the observed pulsar spin-down. 
Besides, one may
notice, following New et al. \cite{NeCJT95}, that if the 
mean ellipticity of pulsars is taken to be of the order of 
the $\epsilon_{\rm max}$ of millisecond pulsars, i.e. 
$\epsilon \sim 10^{-9}$ (cf. Table~\ref{t:hmax}), then the Crab pulsar 
reveals to be a much worse candidate than PSR B1957+20, as it can be seen by
setting $\epsilon = 10^{-9}$ in Eq.~(\ref{e:h0,num}) for these objects: 
$h_0^{\mbox{\tiny Crab}} \simeq 2\times 10^{-30}$ versus 
$h_0^{\mbox{\tiny 1957+20}}\simeq 1\times 10^{-27}$. 

The detectability of {\em individual} NS
by existing and future gravitational wave
detectors has been discussed by various authors, including 
Schutz \cite{Schut91}, Jotania et al. \cite{JotaD94,JotVD96},
Suzuki \cite{Suzuk95}, 
New et al. \cite{NeCJT95}, and Dhurandhar et al. \cite{DhuBC96}. 
For VIRGO-like instruments, it can be hoped that any pulsar that produces
a GW amplitude on Earth, $h_0$, greater than $10^{-26}$ in the frequency 
bandwidth where the sensitivity of the detector is better than 
$10^{-22}{\ \rm Hz}^{-1/2}$, can be detected with three years of integration
\cite{GourB96}. The last column of Table~\ref{t:hmax} gives the minimum
value of $\epsilon$ required to produce $h_0>10^{-26}$.  Notice that for 
Crab-like pulsars 
this value is below $1\%$ of the maximum ellipticity allowed by the 
measured spin-down rate of the pulsar, whereas for millisecond pulsars
both values are of the same order. 

In the present article we study the possibility
to detect the total CW emission from {\em the whole population} 
of NS in our Galaxy
by using only one LIGO/VIRGO type detector.
The proposed strategy consists in measuring the square of the gravitational
signal, $ h^2 $, and in detecting the sidereal modulation of this signal 
which results from the 
directivity of the detector and the anisotropy of the NS distribution.
This technique (referred hereafter as {\em quadratic detection})
is very similar to the one used in radioastronomy 
when only one antenna is used and differs from the strategy proposed by 
Schutz \cite{Schut91}
that consists in searching for NS one by one within a 
4-dimensional space (frequency, phase and position on the sky)
(technique referred hereafter as {\em linear detection}).
The advantages and drawbacks of the two strategies are discussed and it will
be shown that if the number of the emitting stars is larger than 
$2\times 10^6$, then the signal squaring technique is more convenient
and more computer time saving than the linear one.
Actually, the two techniques appear to be complementary. 

The paper is organized as follows:
in Sect.~\ref{s:stat} the statistical properties of the
squared signal are briefly recalled. 
The efficiency of the method as a function of the NS number 
and of the range of frequency at which they are supposed to radiate is computed,
the comparison with the linear analysis for single NS search being
performed in Sect.~\ref{s:compar:linear}. 
In Sect.~\ref{s:stabil}, the constraints on the stability of the
instrumental noise are computed.
In Sect.~\ref{s:galactic}, we try to evaluate the number of gravitationally 
emitting NS in our Galaxy and the range of frequencies at which they are
supposed to radiate. Sect.~VI contains the conclusions. 
\section{Statistical properties of the squared signal} \label{s:stat}
The response $h(t)$ of an interferometric detector to the gravitational 
radiation emitted by
$N$ NS spread out on the sky is detailed in Appendix~\ref{s:append,average}. 
For the purpose of this section, let us consider that $h(t)$ is 
the sum of $N$ elementary periodic functions $h_i$ with unknown frequencies
 $ \nu_i $ and phases $\phi_i$, the precise form being given by
Eqs.~(\ref{e:h(t)})-(\ref{e:Ai}). 

The time average value of $h(t)$, $ \langle h(t)\rangle $, is zero, but the
average of its square is not.
Therefore the strategy to detect the gravitational emission of an ensemble of 
NS, the frequencies of which
spread out from $ \nu_1 $ to $ \nu_2 $, consists in measuring the square
 of the signal $ h^2(t) $. It is easy to see that $\langle h^2(t)\rangle $ is 
proportional to the sum of the squares of the shear from each  
NS (cf.~Eq.~(\ref{e:I(t)=sum})): 
\be \label{e:h2=sum:alpha}
  \langle  h^2(t) \rangle  \simeq   \sum_{i=1}^N  \alpha_i(t) 
	{A_i^2 \ov r_i^2} \ ,
\ee
where  $r_i$ is the distance of the $i$-th NS, $A_i$ its gravitational wave
amplitude at one unit distance, $\alpha_i(t)$ some factor involving
the direction of the NS and the polarization of its radiation with
respect to the detector arms. For the purpose of the
present discussion, Eq.~(\ref{e:h2=sum:alpha}) can be 
recast in the following approximate form
\be \label{e:h2=K(t)}
 \langle  h^2(t) \rangle  \simeq K(t) \, N \, {A^2 \ov 
	D_{\rm q}^2 } \ ,
\ee   
where $A^2$ is the mean value of $A_i^2$, 
\be
  D_{\rm q} := \l[ {1\ov N} \sum_{i=1}^N {1\ov r_i^2} \r] ^{-1/2}
\ee
is the inverse-square average distance of the NS population
and $K(t)$ is a time varying factor, with the periodicity
of one sidereal day. Its non-constancy is induced by the directivity
of the detector and the anisotropy of the NS distribution. 

The study of the efficiency of the quadratic technique is made
easy by the close analogy with the radioastronomy 
observation  technique. Radioastronomers 
measure the square of the electromagnetic field emitted by
a large ensemble of collective modes of a plasma. They are confronted to
the problem of extracting some signal from the noise of the 
receiver.
One possible technique consists in scanning the sky around the source, 
and in measuring
the differences of the total noise on/off of the source. In the case of
gravitational waves, the source is scanned by the
interferometric detector via the rotation of the Earth.

Let us also notice that the problem of detection of the gravitational  
wave emission from an ensemble of NS is very close to the search for 
a gravitational cosmological background discussed in great 
details by Flanagan \cite{Flana93} (see also \cite{Allen96}). 
The main and basic difference is the daily modulation of the signal from the
NS ensemble, which allows the 
search to be performed with {\em only one} detector instead of two
detectors required for the cosmological background detection 
\cite{Flana93}. 

Let $ s(t) = h(t) + n(t) $ be the output of the detector,
$h(t)$ being the
gravitational radiation signal  and $ n(t) $  the detector's
noise. The frequency bandwidth of the detector is supposed to be $ \Delta \nu =
\nu_2-\nu_1 $. By squaring and by averaging $ s(t) $ we obtain:
\be
\langle s^2(t)\rangle  = \langle h^2(t)\rangle  + \langle n^2(t)\rangle 
\ee
because noise and signal are not correlated and consequently the average of
the cross product $ \langle 2n(t)h(t)\rangle  $ vanishes.     
Here the average must be understood as the average on the outputs of
an infinite number of detectors.

If the noise is stationary, $ \langle n^2(t)\rangle  $ is a constant, and 
its value is  
\be
\langle n^2(t)\rangle  = \int_{\nu_1}^{\nu_2} G(\nu) \hat{n}^2(\nu) d\nu 
\ee
 where $ \hat{n}^2(\nu) $ and
$ G(\nu) $ are respectively the noise power per Hz and the frequency
response of the detector. For simplicity we shall assume the noise being
white ($\hat n$ independent of $ \nu $ and $G(\nu)=1$ for 
$ \nu_1 \leq \nu \leq \nu_2\  $). The above
expression then reads 
\be
\langle n^2(t)\rangle =\hat{n}^2 \Delta \nu
\ee
The quantity $ \langle h^2(t) \rangle  $ is time dependent: there are high 
frequency
time variations with a typical frequency of NS rotation, and a
low frequency time variation due to the slow change of the orientation of
the  interferometer arms with respect to the Galaxy induced by the Earth
rotation. This latter frequency, $\nu_{\rm sid}$, is equal to the inverse 
of one sidereal day: 
$ 1/\nu_{\rm sid}=86164.092055 $ s. It is this modulation that 
must be searched for.

In practice, observations are performed with only one detector;
consequently the ensemble average $\langle  \rangle  $ must be replaced by 
an average on time. In what follows we shall consider a simplified case
in order to allow the reader who is not familiar with the radioastronomy
technique, to understand the basic ideas. The reader will find more details
in the excellent book by Kraus \cite{Kraus66} and in the already quoted
paper by Flanagan \cite{Flana93}. Let $T$ be the averaging time. We can write
\be \label{quadn}
{1\ov T} \int_{t}^{t+T} n^2(t')\,  dt' = \langle n^2(t)\rangle  + \psi(t) \ ,
\ee
where $ \psi(t)$ is a random function. Under the ergodicity hypothesis 
$ \psi(t) $ vanishes when $ T \rightarrow \infty $.
When $T$ is finite, $ \psi(t) $ has the following
statistical properties:
\be \label{stpsi}
 \langle \psi(t)\rangle  =0\ ,\quad   \langle \psi^2(t)\rangle  =
	 C \hat{n}^4\Delta\nu/T
\ee
Here $C$ is a constant of the order of unity and depends
on $ G(\nu) $. For the rectangular filter considered above $ C=2$.
Values of $C$ for different filters can be found in Chapter 7 of
ref.~\cite{Kraus66}. 
 
If $ T $ is much shorter than one sidereal day, $\langle  h^2(t) \rangle  $ 
can be considered
as constant, and the signal-to-noise ratio  $ R_{\rm quad} $ can be easily 
computed from Eqs.~(\ref{e:h2=K(t)}) and (\ref{quadn}):
\be
R_{\rm quad}={N \, A^2\ov  D_{\rm q}^2 \sqrt{\langle \psi^2(t)\rangle }} =
{N\, A^2\sqrt{T} \ov D_{\rm q}^2 \ \hat{n}^2\sqrt{C\Delta \nu} }
\ee
If $H = N/\Delta\nu $ is the number of the NS per unit frequency, 
$ R_{\rm quad} $ reads
\be
R_{\rm quad}= {H\, A^2 \sqrt{\Delta\nu T}\ov \hat{n}^2 D_{\rm q}^2 \sqrt{C} }   
\ee
 Note that $ R_{\rm quad} $ is magnified by the factor $ \sqrt{\Delta\nu T} $ well 
known by radioastronomers.

If $T$ is much longer than one sidereal day, we have to take into account the
periodicity of the signal. Let us suppose for simplicity\footnote{The actual 
time
variation of $\langle h^2(t)\rangle $ is given in 
Appendix~\ref{s:append,average}.} 
that the factor $K(t)$ appearing in Eq.~(\ref{e:h2=K(t)})
is harmonic with period $ P_{\rm sid} $ 
equal to one sidereal day and with a known  phase:
 $ \langle h^2(t)\rangle = N\, A^2 \cos(2\pi t/P_{\rm sid}) / D_{\rm q}^2$.
In this case the best way to proceed is to make a Fourier transform of the
squared signal.
The signal-to-noise ratio $ R_{\rm quad} $ is then:
\be \label{e:Rq:delta_nu}
R_{\rm quad}={N A^2 \ov D_{\rm q}^2 \sqrt{\hat{\psi}^2(\nu)\, \delta\nu } }
\ee
where $\delta\nu = 2/T $ is the width of a single bin of the Fourier transform
and $ \hat{\psi}^2(\nu) $ is the spectral power of the noise $\psi(t)^2$.
At very low frequency  $\hat\psi^2(\nu)= C \hat n^4\Delta\nu $, so that
Eq.~(\ref{e:Rq:delta_nu}) becomes
\be \label{Rquad}
  R_{\rm quad}= N \ {A^2 \ov D_{\rm q}^2}\ 
	\sqrt{T\ov 2 C\, \Delta\nu} \, {1\ov {\hat n}^2 } \ . 
\ee
In the realistic case (cf. Appendix~\ref{s:append,average}), 
the signal is periodic (period $P_{\rm sid}$) 
but not harmonic. Its shape depends on the Galactic
NS distribution. Figures~\ref{f:disq} and \ref{f:halo}
show its variation during one
sidereal day for two different NS distributions. 
These shapes can be used as templates to optimize the extraction
of the signal.

\section{Comparison with the linear search for a single NS} 
\label{s:compar:linear}

Let us compare the efficiency of the
quadratic analysis with respect to the linear one  proposed by Schutz 
\cite{Schut91} for single NS searches.
If the frequency and the position of the NS one searches for is known, the 
signal-to-noise ratio of the linear technique is given by
\be \label{Rlin}
R_{\rm lin} = {A \ov D_{\rm i}} \, \sqrt{T\ov 4 {\hat n}^2} \ ,  
\ee 
where $D_{\rm i}$ is the distance to the individual NS that is searched for. 
The factor $4$ in the denominator is due to the product of the 
bandwidth times an extra-factor $2$ coming from the fact that the phase is not
known.

The ratio $ E =R_{\rm lin}/\sqrt{R_{\rm quad}} $ is a good quantity to
 characterize the advantages and drawbacks of the two methods. We have, 
from Eqs.~(\ref{Rquad}) and (\ref{Rlin}),
\be \label{Effic.}
E = {D_{\rm q}\ov D_{\rm i}} \, 
	{(C \Delta\nu T)^{1/4} \ov 2^{3/4} \, N^{1/2} } \ .
\ee
Taking $\Delta\nu = 1{\rm\ kHz}$, $T= 1 {\ \rm yr}$, $C=2$ and
$D_{\rm i} = D_{\rm q}$, 
the quadratic technique appears to be more efficient ($E\leq 1$)
than the linear one if the number of NS is larger than $ 9\times 10^4 $.
The above result is based on the two underlying hypothesis:
(i) the frequency and the position of the isolated NS are known exactly, 
(ii) the distance $ D_{\rm i} $ of the isolated NS is equal to the averaged
distance $D_{\rm q} = \langle  1/r^2_i \rangle ^{-1/2} $ of the NS of the 
ensemble.

Let us consider now the case $D_{\rm q} \not = D_{\rm i}$. 
$D_{\rm q} $ depends on the distribution of  NS in the Galaxy. If their 
distribution corresponds to the Galactic disk (see Sect.~\ref{s:galactic}),
then  $D_{\rm q} = 5.1$ kpc (value resulting from the distribution
function (\ref{e:distrib:disq}) with the parameters $R_0 = 3.8 {\rm\ kpc}$
and $z_0 = 0.5{\rm\ kpc}$). 
$D_{\rm i} $ can be taken to be the distance of the closest NS.
For example, for the nearby millisecond pulsar, PSR J0437-4715 \cite{Johns93},
$D_{\rm i} = 140{\rm\ pc}$. Let us take $D_{\rm i}= 100 {\rm\ pc}$. 
The two methods become then equivalent for $N=2.2\times 10^8$ NS.

However, in searching for a single NS, its position has to be known 
with an accuracy high enough
to compensate for the Doppler shift of the frequency induced by the
rotation of the Earth and its motion 
around the Sun. The last effect is the most important one.
The accuracy of the declination $\delta $ and in the azimuth $\phi $ is about
 $\delta \phi=\phi R_\oplus \nu_{\rm max} c$ where 
 $R_\oplus$ , $\nu_{\rm max}$ and $c $ are respectively the
radius of the Earth orbit ($150\times 10^6 {\rm\ km}$), 
the expected maximum frequency 
of the NS and the speed of 
light. A similar precision is needed on the declination of the source. 
This means that the sky must divided in about 
$4\times 10^{10}\ (\nu/100 {\ \rm Hz})^2 $ boxes
and we have to try to detect a periodic source (by Fourier transform)
in each box, after compensation of the variation of the frequency of the source
due to the Earth motion. In the above rough analysis, we have neglected the
less important Doppler shift induced by the Earth rotation.
In this section, we do not discuss the technical possibility of performing
more than $10^{10} $ Fourier transforms each of them containing 
$ 3\times 10^9(\nu_{\rm max} /100 {\ \rm Hz}) (T / 1{\ \rm yr}) $ bins.
The detection will be considered as positive if the probability  of a 
random fluctuation of the signal in {\em all } the  boxes times the
number of bins is less than a prefixed value. If we take the probability to be 
lower than .16 (corresponding to the  1 $\sigma$ criterion), the value of the 
corresponding signal-to-noise ratio $R_{\rm lin}$ is about 9.5 for 
$\nu_{\rm max}=100{\rm\ Hz} $ 
and 10.5 for $\nu_{\rm max}= 1 {\rm\ kHz}$. 
Consequently Eq.~(\ref{Effic.}) becomes
\be \label{Efic2}
E=10\times {D_{\rm q}\ov D_{\rm i}} \, 
	{ (C \Delta\nu T)^{1/4}\ov 2^{3/4} N^{1/2}}
\ee
The two methods become then equivalent for $ N = 2.2\times 10^6 $.

\section{Stability of the noise} \label{s:stabil}

\subsection{Non-stationary noise}

The above results hold under the hypothesis that the term 
$\langle n^2(t)\rangle  $ in Eq.~(\ref{quadn})
 is  constant, i.e. that the noise is
stationary. Now the actual noise is not stationary: low frequency 
fluctuations are always present. Let us recall that the sources of noise 
in interferometric
detectors are the photon shot noise (at high frequency) and the Brownian noise
of the mirrors (at low frequency).
There are at least two sources of low frequency fluctuations: (i)
the fluctuations of the optical power of the laser and (ii) 
the temperature fluctuations of the mirrors. 
When the interferometer is in lock, the
fluctuations of the laser power can be taken under control within few 
$10^{-5}$ and are therefore not dangerous at all.
However, the interferometer may be falling out of lock occasionally
or the laser may be shut down and switched on some time later. In either
case, this will induce a temperature fluctuation in the mirrors. However its
amplitude will be at most $0.1{\ \rm K}$ \cite{Vinet96}. This falls within the
range of the required accuracy on the temperature measurement of non-periodic
fluctuations, as derived in Sect.~\ref{s:non-periodic} below. 

We shall thus 
consider only the mirror temperature fluctuations
and estimate the constraints they impose. 
We will discuss the possibility of monitoring 
the temperature fluctuation to keep their effects under control.

In what follows, we suppose that the fluctuations
in the term $\langle n^2(t)\rangle $ of  Eq.~(\ref{quadn}) have a
small amplitude and that their typical time scale
is much longer than $1/\nu_1$. Let us introduced the new random 
variable $\alpha(t)$ by
\be \label{nsta}
\langle n^2(t)\rangle  = \overline{\langle n^2(t)\rangle }(1 + \alpha(t))   
\ee
where the symbol $\overline{X}$ means the average on time (if it exists).
The hypothesis of small amplitude fluctuations implies that 
$\sqrt{\langle \alpha^2(t)\rangle }\ \ll 1$. 
As stated above, we consider only the contribution to the noise 
due to the temperature. 
In this case, the noise ${\hat n}^2(\nu)$ is proportional to the
temperature $\Theta$ \cite{Sauls90} \cite{BondV95} : 
\be \label{e:hn2=bt}
  {\hat n}^2(\nu) = \beta \Theta \ .
\ee
Under the above hypotheses, Eq.~(\ref{quadn}) reads:
\be \label{quadn2}
  {1\ov T} \int_{t}^{t+T} n^2(t') \, dt' = 
	\overline{\langle n^2(t)\rangle }\, (1+\alpha(t)) 
	+ \psi(t)
\ee
where $\psi(t)$ has the same properties as that stated in Eq.~(\ref{stpsi}). 
 Within quadratic terms in $\alpha(t)$,  Eq.~(\ref{quadn2}) reads
\be \label{quadn3}
{1\ov T} \int_{t}^{t+T} n^2(t')\,  dt' = 
  \hat{n}^2 \, \Delta\nu + \hat{n}^2 \, \Delta\nu\, \alpha(t)
 + \psi(t) \ .
\ee
The random variable $\alpha(t)$  has a zero mean value,
$ \overline{\alpha(t)}=0$, and  its variance $\overline{\alpha^2(t)} $ 
is proportional to the temperature fluctuations of the mirrors. Consequently
the typical
time scale for $\alpha(t)$ is of the order of one hour, much longer than 
$1/\nu_1 \sim 0.1 {\rm\ s}$. The typical amplitude of the temperature variation
is a few kelvins, much smaller than the room temperature ($300 {\rm \ K}$),
therefore the above hypotheses are fulfilled.
It is worth to note that in defining $\alpha(t)$ and its statistical properties
we have averaged on time, instead of averaging on an infinite ensemble of 
identical detectors. The reason is that for {\em identical} detectors in the
same environment, the daily variation of the temperature, and
consequently the amplitude of the thermal noise, is the same for {\it all}
the detectors.

\subsection{Non-periodic temperature fluctuations} \label{s:non-periodic}

The fluctuations of the mirrors temperature are not known today. 
Only in situ measurements of the temperature, once the detector 
will be operational, will provide us with 
the statistical properties of $\alpha(t)$. 
However, let us try to guess the temperature trends of the mirrors, 
in order to see if the precision needed
on the temperature control can be achieved with the present technology.
 One reasonable hypothesis is to take a power law for the Fourier spectrum of
 the temperature fluctuations $\delta\hat\Theta(\nu)^2$ :
\be \label{spectrum}
\delta\hat\Theta(\nu)^2 = {a^2\ov \nu_{\rm T}} 
	\l( {\nu\ov \nu_{\rm T}} \r) ^{-\gamma} 
\ee
where $ a $, $\gamma$ and $\nu_{\rm T}$ are some constants. 
These parameter can be estimated in the following way. 
According to Eq.~(\ref{spectrum}), the temperature variation on a time scale 
$\tau$ is given by
\be \label{e:dtheta2}
   \langle \delta \Theta^2\rangle  =
	{a^2\ov \nu_{\rm T}} \int_{\nu}^{\infty}
	\l( {\nu'\ov \nu_{\rm T}} \r) ^{-\gamma}\, d\nu'=
	{a^2\ov \gamma-1} \l( {\nu\ov \nu_{\rm T}} \r) ^{1-\gamma} \ , 
\ee
where $\nu= 1/\tau$. 
As already said, we do not have yet any real measurements of the temperature
variations of the detector's mirrors. Consequently we shall proceed by guessing
the temperature variations in the environment of the experiment. It seems 
reasonable to assume that the temperature variation on a time scale 
$\tau = 1 {\rm\ h}$ cannot exceed $1{\rm\ K} $ and $10{\rm\ K}$ 
on a time scale $\tau = 12 {\rm\ h}$. Setting 
$\nu_{\rm T} = (1 {\rm\ h})^{-1} = 1/3600{\rm\ Hz}$ in Eq.~(\ref{e:dtheta2}),
we obtain then $\gamma = 3$ and $a=1.41\ {\rm\ K}$.

Let us now estimate the precision required in the temperature monitoring 
in order that the amplitude of the temperature noise be lower than the
``ordinary'' noise $\psi(t)$.  From Eq.~(\ref{quadn3}), this requirement
writes $\hat{n}^4 \, \Delta\nu^2 \hat \alpha(\nu)^2 < \hat\psi(\nu)^2$.  
Taking into account that $\hat \alpha(\nu)^2 = \delta\hat\Theta(\nu)^2 /
\Theta^2$, $\hat\psi^2(\nu)=C\hat{n}^4\Delta \nu$
(Eq.~(\ref{stpsi})), and
$\gamma = 3$,  one obtains from Eq.~(\ref{spectrum}) the condition (for $C=2$):
\be
   a < \sqrt{2 }\, \Theta \, {\nu^{3/2}\ov \nu_{\rm T} \, \Delta\nu^{1/2} }
   = 4.8 \times 10^{-3}  \left( \frac{\Theta}{300{\rm\ K}} \right)
   \left( {\nu \ov 10^{-5}{\rm\ Hz} } \right) ^{3/2}  
    \left( \frac{100{\rm\ Hz}}{\Delta \nu}\right) ^{1/2}\ {\rm\ K} \ .
\ee
For $\nu=(12{\rm\ h})^{-1}$, $\Delta\nu=100 {\rm\ Hz}$, and
$\Theta=300{\rm\ K}$, we get $a<1.7\times 10^{-2}{\rm\ K}$. This value
is one hundred times smaller than the actual value of $a$ obtained above
from the expected daily temperature variation ($a=1.41\ {\rm\ K}$). Since
at this frequency ($(12{\rm\ h})^{-1}$), the amplitude of the temperature
variation is about $10{\rm\ K}$, this means that the temperature fluctuations
have to be monitored with a precision of $10^{-2} \times 10{\rm\ K} = 
0.1{\rm\ K}$ for the extra-noise to be extracted from the ``ordinary''
noise $\psi(t)$.

\subsection{Periodic temperature fluctuations}

A {\em periodic} modulation
due to the variation of the solar flux might be present in the power spectrum
of the mirror temperature fluctuations. 
Maybe this will not be the case, given the amount of shielding the 
vacuum system and suspension will provide to the mirrors. We however consider
temperature fluctuations correlated with the solar flux because they represent
the most dangerous obstacle to the proposed method of detection. Indeed 
the solar flux spectrum contains, beside the solar day frequency, 
the sidereal day one, which may pollute the GW signal from 
galactic NS. 
Table~\ref{tab1} shows the Fourier spectrum of the solar flux 
at the latitude of Pisa (VIRGO site) computed by taking a period of 4 years.
There are important lines at frequencies corresponding to multiples of
the inverse of one solar day 
($\nu_{\rm sol}=1.1574074\times 10^{-5} {\rm\ Hz}$), as well as  
important lines at frequencies that are multiple of the inverse of one
sidereal day ($\nu_{\rm sid}= 1.1605763\times 10^{-5}{\rm\ Hz}$).  
These latter lines can be explained by looking at the expression giving 
the solar flux as a function of time: the daily solar flux is modulated by the 
annual variation of the declination of the Sun. Consequently, there exists
a line resulting from the beating of the solar frequency
$\nu_{\rm sol} $ with the frequency corresponding to the inverse of 
the tropical year ($ \nu_{\rm ty}=3.1688765\times 10^{-8}{\rm\ Hz}$). 
Therefore around the solar frequency, the 
two frequencies $\nu_{\rm sol}\pm\nu_{\rm ty}$ are present, and in 
particular the sidereal day frequency 
$\nu_{\rm sid}=\nu_{\rm sol}+\nu_{\rm ty} $.

The possible temperature fluctuation spectrum induced by the solar flux 
can be written  
\be \label{spec2}
  \delta\hat\Theta(\nu)^2 = 
 \sum_j b^2_j\delta(\nu-\nu_j)\ , 
\ee
where the $\nu_j$ are the solar flux frequencies displayed in Table~\ref{tab1}. 

Let us compute the level at which the temperature of the mirrors 
has to be taken under control, in order to have the amplitude of this
extra-noise lower than the ``ordinary'' noise $\psi(t)$.
Taking into account Eq.~(\ref{quadn3}), the relationship between
the coefficients $b_j$ (Eq.~(\ref{spec2})) and the periodic components
$\alpha_j$ of the noise $\alpha(t)$ reads
\be
  \langle  b_j^2 \rangle =  
   \beta^2 \Theta^2 \, \Delta\nu^2  \langle  \alpha_j \rangle ^2 \ .
\ee
For a given multiple of the sidereal frequency, $j\nu_{\rm sid}$, we have thus
to compare the contribution of the extra noise 
$\alpha_j {\hat n}^2(\nu) \, \Delta\nu \cos(2\pi j \nu_{\rm sid} t +\phi)$
with the term $\psi(t)$ (Eqs.~(\ref{quadn}), (\ref{stpsi}) and 
(\ref{quadn2})). Taking into account that the thickness of the bin of a
Fourier transform of a signal of length $T$ is $2/T$, we have
$\alpha_j \leq \sqrt{2C/\Delta\nu T}$; the corresponding periodic
temperature variation $\sqrt{\langle \delta \Theta^2 \rangle }$  
must be less  than $10^{-2}(\frac{\Delta\nu}{100{\rm\ Hz}})^{-1/2}
T^{-1/2}_{\rm year}{\rm\ K} $ ($C=2$).     

\subsection{Conclusions}

From the above analysis it appears that the non-periodic temperature 
fluctuations of the mirrors are not very dangerous, because an accuracy
of $0.1{\rm\ K}$ in the mirror temperature measurement seems easily
reachable. On the contrary, the periodic 
fluctuations must be kept under control within a few $10^{-3}{\rm\ K}$ for 
observation times longer than one year. 
This seems to be a challenging task. However, it must be noticed that the
periodic temperature fluctuations can be measured on time intervals of
the order of one year, and therefore are easier to control. Note also that
measures of the noise at frequencies $ \nu_{\rm sol}$  and 
$\nu_{\rm sol}-\nu_{\rm ty}$
allows to deduce the noise at the sidereal frequency 
$\nu_{\rm sid}=\nu_{\rm sol}+\nu_{\rm ty}$ for the
spectral components at the frequencies $\nu_{\rm sol} - \nu_{\rm ty}$
and $\nu_{\rm sol} + \nu_{\rm ty}$ are almost identical (cf. Table~\ref{tab1}). 
The required accuracy  of these measurements is about $10^{-3}$.

To conclude, we recommend that all the detector parameters 
(temperature of the mirrors, power of the laser and so on)
shall be monitored. In fact, as already said, we do not need any regulation
of the temperature, but only to know, with the accuracy discussed above, 
{\em how} the temperature varies.

\section{Number of neutron stars and their galactic distribution}
\label{s:galactic}

\subsection{Number of rapidly rotating neutron stars in the Galaxy}
\label{s:numberNS}

\subsubsection{General considerations}

From the star formation rate and the outcome of supernova explosions, 
the total number of NS in the Galaxy is estimated to be about
$10^9$ \cite{TimWW96}. The number of {\em observed} NS is
much lower: $\sim 700$ NS are observed as
radio pulsars \cite{TayML93,TaMLC95}, $\sim 150$ as 
X-ray binaries \cite{WhiNP95,Parad95} (among which $\sim 30$ are X-ray pulsars)
and a few as isolated NS, through 
their X-ray emission \cite{WalWN96}. 

For our purpose, the relevant number is given by the fraction of 
these $\sim 10^9$ NS which rotates sufficiently rapidly to emit
gravitational waves in the
frequency bandwidth of VIRGO-like detectors. 
The upper bound of this bandwidth ($\nu_{\rm max} \sim \mbox{a few kHz}$),
is sufficiently high to encompass even the most rapidly rotating NS, at the
centrifugal break-up limit, which --- depending on the nuclear matter equation
of state and on the NS mass --- ranges from $1 {\rm\ kHz}$ to 
$2 {\rm\ kHz}$ \cite{SaBGH94}. On the other hand, the lower bound of the
interferometer bandwidth ($\nu_{\rm min}\sim 10 {\rm\ Hz}$) is a sensitive
parameter for increasing the number of accessible NS. If the observed 
radio pulsars are representative of the population of rotating NS,  lowering 
$\nu_{\rm min}$ from $10 {\rm\ Hz}$ to $5 {\rm\ Hz}$ would increase the number
of observable NS by a factor $2.3$.

In the following, we set the low
frequency threshold of VIRGO to the value $\nu_{\rm min} = 5{\ \rm Hz}$, 
which may be reachable in a second stage of the experiment. 
Using the fact that the highest gravitational frequency of a
star which rotates at the frequency $\nu$ is $2\nu$, 
this means that the rotation period of a detectable NS must be lower
than $P_{\rm max} = 0.4{\ \rm s}$. 
The NS that satisfy to this criterion can be divided into three 
classes:
\begin{description}
\item[(C1)] young pulsars which are still rapidly rotating 
(e.g. Crab or Vela pulsars);
\item[(C2)] millisecond pulsars, which are thought to have been spun up by
accretion when member of a close binary system (during this phase the system 
may appear as a low-mass X-ray binary); 
\item[(C3)] NS with $P<0.4{\ \rm s}$ but 
which do not exhibit the pulsar phenomenon. 
\end{description}

In the examples given in Table~\ref{t:hmax}, the first three entries
belong to class (C1), while the other two ones belong to class (C2). 

The number of millisecond pulsars in the Galaxy
is estimated to be of the order $N_2 \sim 10^5$ \cite{Bhatt95}. 
The number of {\em observed} millisecond pulsars ($P<10 {\ \rm s}$)
is about 50 and is continuously increasing.   

The number of young (non-recycled) rapidly rotating NS is more difficult
to evaluate. An estimate can be obtained from the fact that the
observed non-millisecond pulsars with $P<0.4{\ \rm s}$ represent
28 \% of the number of cataloged pulsars and that the 
total number of active pulsars in the Galaxy is around $5\times 10^5$
\cite{Lyne95}. The number of rapidly rotating non-recycled pulsars 
obtained in this way is $N_1 \sim 1.4\times 10^5$. 

Adding $N_1$ and $N_2$ gives a number 
of $\sim 2\times 10^5$ NS belonging to the populations (C1) and (C2) 
defined above. 
This number can be considered as a lower bound for the total number
of NS with $P<0.4 {\ \rm s}$. The final figure depends on the
amount of members of the population (C3). This latter number is (almost by
definition !) unknown. We present below a scenario leading to a
large and potentially detectable population (C3). 

\subsubsection{The specific case of NS relics of the Galaxy
formation}

It seems now well
established that the birth of galaxies has been accompanied by a burst of 
massive-star formation. This took place at a cosmological redshift
$z\sim 2$ \cite{LiLHC96}, \cite{MFDGS96}, i.e. when the Universe was 
$\sim 20\%$ of its present age. The remnant of these first generations
of massive stars could contribute significantly to the population (C3), 
as we are going to see. 

The massive stars formed in the infancy of the Galaxy, some 
$\tau \sim 10{\rm\  Gyr}$ ago, should have given birth to a large population
of rapidly rotating ($P^{-1} > 20 {\rm\ Hz}$) NS. 
Let us assume that these NS have a mean ellipticity of 
$\epsilon \simeq 10^{-6}$, which amounts to 
only one thousandth of the maximum ellipticity permitted for 
the Crab pulsar (cf. Table~\ref{t:hmax}). 
Let us consider the fraction of these stars
for which the spin-down is driven by gravitational radiation and not
by electromagnetic processes. This may happen in the following cases:
(i) the magnetic field is quite low ($B<10^{10}{\rm\ G}$) either because
the NS have been formed with such a field --- given our poor knowledge of 
magnetohydrodynamical processes during stellar collapse and the proto-neutron
star phase, this cannot be excluded --- or because it has been 
destroyed during an accretion phase
in a close binary system \cite{UrpiG95}, (ii) the magnetic field configuration
is such that the energy losses are small (for a review of the possible
magnetic field structure of NS and their evolution see e.g. 
ref.~\cite{BhatS95}). 
Under these assumptions, the considered 
NS do not show up at present as radio pulsars, i.e. they belong to
population (C3). Note that New et al. \cite{NeCJT95} have also 
suggested that there could exist a large population of NS whose 
spin-down is driven by gravitational radiation. The differences with
our hypothesis are that (i) they consider these NS to be 
presently rapidly rotating 
(millisecond periods)  
and (ii) they do not provide any specific scenario to create this population. 

An easy calculation show that with $\epsilon \simeq 10^{-6}$ and in 
$10{\rm\ Gyr}$, the emission of gravitational radiation has slow down these
NS to rotational period of $P=0.075{\rm\ s}$, quite independently of their
initial period. This corresponds to a frequency of $13{\rm\ Hz}$, which
fits in the low frequency part of interferometric detectors (contrary
to the population suggested by New et al. \cite{NeCJT95}).  
The present gravitational wave amplitude 
emitted by such a NS at one distance unit is given by Eq.~(\ref{e:Ai}) below
and amounts to $A\simeq 7.5 \times 10^{-28}{\rm\ kpc}^{-1}$. 
Inserting this value into Eq.~(\ref{Rquad}) and taking 
the inverse-square average distance of this NS population to be
$D_{\rm q} = 5{\rm\ kpc}$ (cf. Sect.~\ref{s:compar:linear}) leads to the
following signal-to-noise ratio for the quadratic detection of these NS,
the number of which is $N$:
\be
  R_{\rm quad} = 0.34\ \l( {N \ov 10^{10}} \r) 
	\l( {T\ov 3{\rm\ yr}} \r)
	\l( {10 {\rm\ Hz}\ov \Delta\nu} \r) ^{1/2}
	\l( {10^{-21} {\rm\ Hz}^{-1/2} \ov \hat n} \r) ^2 \ ,
\ee
where $10^{-21} {\rm\ Hz}^{-1/2}$ is the (present day) expected 
VIRGO sensitivity at the frequency of $10{\rm\ Hz}$ \cite{Giazo95}
and $10^{10}$ is a possible value for the total number of relic NS. 
Note that since the considered NS population is supposed to radiate
at low frequencies, one can take a narrow bandwidth $\Delta\nu = 10{\rm\ Hz}$ in
order to increase the signal-to-noise ratio. Note also that improving
the low-frequency detector sensitivity at from $10^{-21} {\rm\ Hz}^{-1/2}$
to say $10^{-22} {\rm\ Hz}^{-1/2}$, would lead to a signal-to-noise
ratio of $\sim 30$ if there are $N\sim 10^{10}$ relic NS with an
mean ellipticity of $10^{-6}$.

\subsection{Galactic distribution}

The {\em observed} pulsars are concentrated toward the galactic plane,
with a scale height above that plane
of about $0.5{\ \rm kpc}$ \cite{Lyne95}. This latter 
value is almost an order of magnitude greater than the scale height of
their progenitors  (massive stars), reflecting the high ``kick'' velocities
that pulsars acquire at their birth (see e.g. \cite{BurrH96} and
references therein). If the rapidly rotating NS considered in 
Sect.~\ref{s:numberNS} follow this distribution, their distribution
function can be represented by Eq.~(\ref{e:distrib:disq}) and
the corresponding time variation of the squared GW signal 
$\langle  h^2(t) \rangle $
is shown in Fig.~\ref{f:disq}. The Fourier spectrum of this signal is
given in Table~\ref{t:fourier:h2}. Beside the constant part,
it involves four frequencies, which are the first four harmonics of the
sidereal day frequency $\nu_{\rm sid}$. The amplitude of the GW signal
read in Fig.~\ref{f:disq} is 
\be
	\sqrt{ \langle  h^2(t) \rangle } \simeq 2\times 10^{-26}	
		\sqrt{ N \ov 10^5 } \ .
\ee
This value can be compared with the GW amplitude of an individual NS
which would have the same parameters as the mean ones used in the
computation leading to Fig.~\ref{f:disq}, namely $\epsilon = 10^{-8}$,
$P=5 {\ \rm ms}$ and $I=10^{38} {\ \rm kg\, m}^2$ [cf. Eq.~(\ref{e:h0,num})]
\be
	h_0 \simeq 2 \times 10^{-27} \ \l( {{\rm kpc}\ov r} \r) \ .
\ee

Because of the important kick velocities mentioned above, it cannot
be excluded that the actual distribution of NS constitutes a halo around
our Galaxy, instead of being concentrated in the galactic disk. 
A corresponding distribution function if then given by 
Eq.~(\ref{e:distrib:halo}) and the resulting signal is shown in 
Fig.~\ref{f:halo}.
 
From Figs.~\ref{f:disq} and \ref{f:halo}, it appears that, from the 
detection point of view, the disk
distribution is more favorable than the halo one. Indeed, the relative
variation of $\langle h^2(t) \rangle$ during one day is $100\%$ in the
case of the disk, but only $\sim 15\%$ for the halo. This is
due to the fact that the halo distribution is more isotropic than the disk
one.

\section{Conclusion}

The detection of the gravitational waves emitted by rotating NS is a 
difficult task, but a positive result will give important informations
on the evolution of massive stars. Two strategies are conceivable:
the detection of the coherent signal from some isolated NS 
(linear detection) and the detection
of the incoherent signal emitted by a large ensemble of NS 
(quadratic detection).
The two strategies are complementary. The first one is well suited to detect
the gravitational radiation emitted by a few very close NS, 
whereas the second one
allows us to detect the gravitational radiation emitted by {\em all} the NS
of the Galaxy. We have shown that if the distance of the closest NS
is about $100{\rm\ pc}$, 
the two methods give the same probability of detection if the
number of radiating NS in the Galaxy is about $10^6$.

The two strategies have their own drawbacks. Searching for individual radiating
NS requires an extraordinary amount of computational time. It is possible
that the overall sensitivity of this method will be limited by the 
power of future computers. The second method needs a high noise stability
if it is to be used by  {\em  only one detector}. Moreover,
this method has more chances to succeed if the number of emitters in the 
frequency bandwidth of the detector is large, which motivates the 
attempts to keep the low frequency threshold of interferometric detectors as
low as possible. Besides, we have argued that a possible first generation
of NS originating from an important 
stellar formation rate at the birth of the Galaxy, should have been spun down
in $10{\rm\ Gyr}$ 
to periods of the order $0.1{\rm\ s}$ by the gravitational radiation reaction.
This population could be detected by the quadratic method provided that
the sensitivity of interferometric detectors is sufficiently good 
around $5$ to $10{\rm\ Hz}$. It should be noticed that since the wavelengths
corresponding to these frequencies are between $3\times 10^4{\rm\ km}$ and
$6\times 10^4{\rm\ km}$, detectors at different locations onto the Earth 
can be employed for a search in cross-correlation. 

Finally, let us note that the quadratic technique presented in this article
can also be applied to detect a single strong source of
continuous wave gravitational radiation. 

\acknowledgments

We thank Riccardo Mannella, 
Jean-Yves Vinet and Olivier Le~F\`evre for very fruitful discussions. 

\appendix

\section{Squared signal from the NS ensemble} \label{s:append,average}

The detector output is
\be
   s(t) = h(t) + n(t) \ ,
\ee
where 
$n(t)$ is the detector's noise and $h(t)$ the detector's response to 
the gravitational radiation from the $N$ galactic NS:
\be \label{e:h(t)}
   h(t) = \sum_{i=1}^N \l[ F_+^i(t) h_+^i(t)
	+ F_\times^i(t) h_\times^i(t)  \r] \ . 
\ee
In the above equation, $h_+^i(t)$ and $h_\times^i(t)$ are the two polarization
modes of the gravitational waves emitted by the $i$-th NS and
$F_+^i(t)$ and $F_\times^i(t)$ are the (time-varying) 
beam-pattern factors taking into account 
the direction and polarization of the radiation from the $i$-th NS
with respect to the detector's arms (cf. Sect.~5.1 of ref.~\cite{BonaG96}). 
$h_+^i$ and $h_\times^i$ can be expressed in terms of the 
the $i$-th NS's angular velocity $\Omega_i$, ellipticity $\epsilon_i$, 
distance $r_i$, inclination of the rotation axis with respect to 
the line of sight $\iota_i$ and distortion angle $\chi_i$, according to
(cf. Eqs.~(20), (21) and (25) of ref.~\cite{BonaG96})
\begin{eqnarray}
   h_+^i(t) & = & {A_i\ov r_i}  \sin\chi_i \Big[
	{1\ov 2} \cos\chi_i\sin \iota_i\cos \iota_i 
	\cos(\Omega_i t + \phi_i ) \nonumber \\
	& & \qquad \qquad
 - \sin\chi_i {1+\cos^2 \iota_i\ov 2} \cos2(\Omega_i t+\phi_i) \Big] 
						\label{e:h+,gen} \\
   h_\times^i(t) & = & {A_i\ov r_i} \sin\chi_i \Big[
	{1\ov 2} \cos\chi_i\sin \iota_i \sin(\Omega_i t + \phi_i) \nonumber \\
	& & \qquad \qquad
	- \sin\chi_i \cos \iota_i \sin2(\Omega_i t + \phi_i) \Big]
			 \ , \label{e:hx,gen}
\end{eqnarray}
\be \label{e:Ai}
    A_i = {16\pi^2 G\ov c^4} {I_i\, \epsilon_i\ov P_i^2} \ , 
\ee
where $P_i = 2\pi /\Omega_i$ is the rotation period of the star, $I_i$ its
moment of inertia, and $\phi_i$ some phase angle. 

Let us consider the mean value 
\be
    \langle  h^2(t) \rangle  := {1\ov \tau} \int_{t}^{t+\tau} h^2(t') \ dt' 
\ee
of $h^2(t)$ on a time $\tau$ such that
\be \label{e:tau}
   \Omega^{-1}  \ll \tau \ll 1 {\rm\ day} \ ,
\ee
where $\Omega$ is a typical NS rotation frequency: $\Omega^{-1} < 1{\rm\  s}$ 
(for instance $\tau = 10{\rm\ s}$). 
From Eqs.~(\ref{e:h+,gen})-(\ref{e:hx,gen}), and 
using (\ref{e:tau}) ($\tau \gg \Omega^{-1}$ implies that
integrals of products like
$\cos(\Omega_i t)\times \cos(\Omega_j t)$ are negligibly small when
$i\not = j$ and $\tau \ll 1 {\rm\ day}$ implies that $F_+^i$ and $F_\times^i$
are approximatively constant on a time $\tau$) one obtains
\be
   \langle  h^2(t) \rangle  \simeq I_1(t) + I_2(t) + I_3(t) 
\ee
with
\begin{eqnarray}
  I_1(t) & = & \sum_{i=1}^N  {(F_+^i)^2\ov \tau} \int_t^{t+\tau} 
			(h_+^i)^2\ dt'		\\
  I_2(t) & = & 2 \sum_{i=1}^N  {F_+^i \, F_\times^i \ov \tau} \int_t^{t+\tau} 
			h_+^i\, h_\times^i \ dt'		\\
  I_3(t) & = & \sum_{i=1}^N  {(F_\times^i)^2\ov \tau} \int_t^{t+\tau} 
			(h_\times^i)^2\ dt'		
\end{eqnarray}
It can be seen easily that $I_1(t)=I_3(t)$ and,
by virtue of (\ref{e:tau}), $|I_2| \ll I_1$. Hence
\be
  \langle  h^2(t) \rangle  = 2 \sum_{i=1}^N  {[F_+^i(t)]^2\ov \tau} 
	\int_t^{t+\tau} 		[h_+^i(t')]^2\ dt'		\ . 
\ee
Using Eq.~(\ref{e:h+,gen}) for $h_+^i$ and performing the time integration
leads to 
\begin{equation}
   \langle  h^2(t) \rangle   =  {1\ov 4} \sum_{i=1}^N {A_i^2\ov r_i^2} 
	\, [F_+^i(t)]^2 \,
	\sin^2\chi_i \Big[ \cos^2\chi_i \sin^2\iota_i \cos^2\iota_i 
   + \sin^2\chi_i (1+\cos^2\iota_i)^2 \Big] \ . \label{e:I(t)=sum}
\end{equation}

For a given NS distribution, the expression (\ref{e:I(t)=sum}) can
be computed step by step, using the property that
for two {\em independent}
random variables $u$ and $v$, and for large $N$, 
$\sum_{i=1}^N u_i v_i \simeq \bar u \sum_{i=1}^N v_i$, where
$\bar u$ denotes the mean value of $u$: 
$\bar u \simeq  \sum_{i=1}^N u_i\ /\ N$. The variables $A$,
$\chi$ and $\iota$ being independent, 
Eq.~(\ref{e:I(t)=sum}) results in 
\begin{equation}
   \langle  h^2(t) \rangle   =  {\overline{A^2} \ov 4}  
   \Big[	\overline{ \sin^2\chi  \cos^2\chi}\  
 	\overline{ \sin^2\iota \cos^2\iota }
	+ \overline{\sin^4\chi}\ 
       \overline{(1+\cos^2\iota_i)^2 } \Big] \ 
	\sum_{i=1}^N  { [F_+^i(t)]^2 \ov r_i^2} \ .
\end{equation}
Taking  for $\chi$ and $\iota$ distributions that correspond
respectively to the probability law $P(\chi) = 1/\pi$ (uniform distribution 
in $[0,\pi]$) and $P(\iota) = 1/2\, \sin\iota$ (uniform distribution on 
the celestial sphere) results in 
\be \label{e:I(t)2}
    \langle  h^2(t) \rangle  = {43 \overline{A^2} \ov 240} 
   \ \sum_{i=1}^N  { [F_+^i(t)]^2 \ov r_i^2} \ .
\ee
$F_+^i(t)$ depends on (i) the orientation of the detector's arms, 
which varies with time due to the Earth motion,
(ii) the direction of the $i$-th NS, which can be 
represented by its equatorial coordinates on the celestial sphere,
namely the right ascension $\alpha_i$ and the declination 
$ \delta_i$ and (iii) the polarization angle $\psi_i$ of the gravitational
wave with respect to the equatorial coordinates:
\be
	F_+^i(t) = F_+(\alpha_i,\delta_i,\psi_i,t) \ .
\ee
The exact dependence 
is quite complicated and can be found in Sect.~5.1 of ref.~\cite{BonaG96}. 
Let us first take the mean value of $F_+^i(t)$ with respect to the 
polarization angle $\psi$, which is uniformly distributed in $[0,2\pi]$. 
Let us denote the result by $\overline{{}^\psi F}_+$:
\be
   \overline{{}^\psi F}_+ (\alpha_i, \delta_i,t) = 
	\langle F_+(\alpha_i,\delta_i,\psi,t) \rangle _{\psi} \ .
\ee
Equation~(\ref{e:I(t)2}) then becomes
\be \label{e:I(t)3}
    \langle  h^2(t) \rangle  = {43 \overline{A^2} \ov 240} 
   \ \sum_{i=1}^N  
   {\l[ \overline{{}^\psi F}_+ (\alpha_i, \delta_i,t) \r] ^2
 	\ov r_i^2} \ .
\ee

Let $f(r,\alpha,\delta)$ be the spatial distribution function of 
NS in the Galaxy, normalized so that $N\,f(r,\alpha,\delta)$
is the number density of NS: 
\be
    \int_{r=0}^{r=\infty} \int_{\delta=-\pi/2}^{\delta=\pi/2}
	\int_{\alpha=0}^{\alpha=2\pi}  f(r,\alpha,\delta) 
	\, r^2\, \cos\delta \, dr\, d\delta\, d\alpha \ = 1 \ .
\ee
Equation~(\ref{e:I(t)3}) becomes 
\be \label{e:I(t):final}
     \langle  h^2(t) \rangle  = {688\ov 15} {\pi^4 G^2\ov c^8} \, 
	\overline{I^2}\ 
	 \overline{\epsilon^2} \  \overline{P^{-4}} \ N \  
     \int_{r=0}^{r=\infty} \int_{\delta=-\pi/2}^{\delta=\pi/2}
	\int_{\alpha=0}^{\alpha=2\pi} 
   \l[ \overline{{}^\psi F}_+ (\alpha, \delta,t) \r] ^2
	\, f(r,\alpha,\delta) \, \cos\delta \, dr\, d\delta\, d\alpha \ ,
\ee
where use has been made of Eq.~(\ref{e:Ai}) to express $\overline{A^2}$. 

\begin{figure}
\centerline{ \epsfig{figure=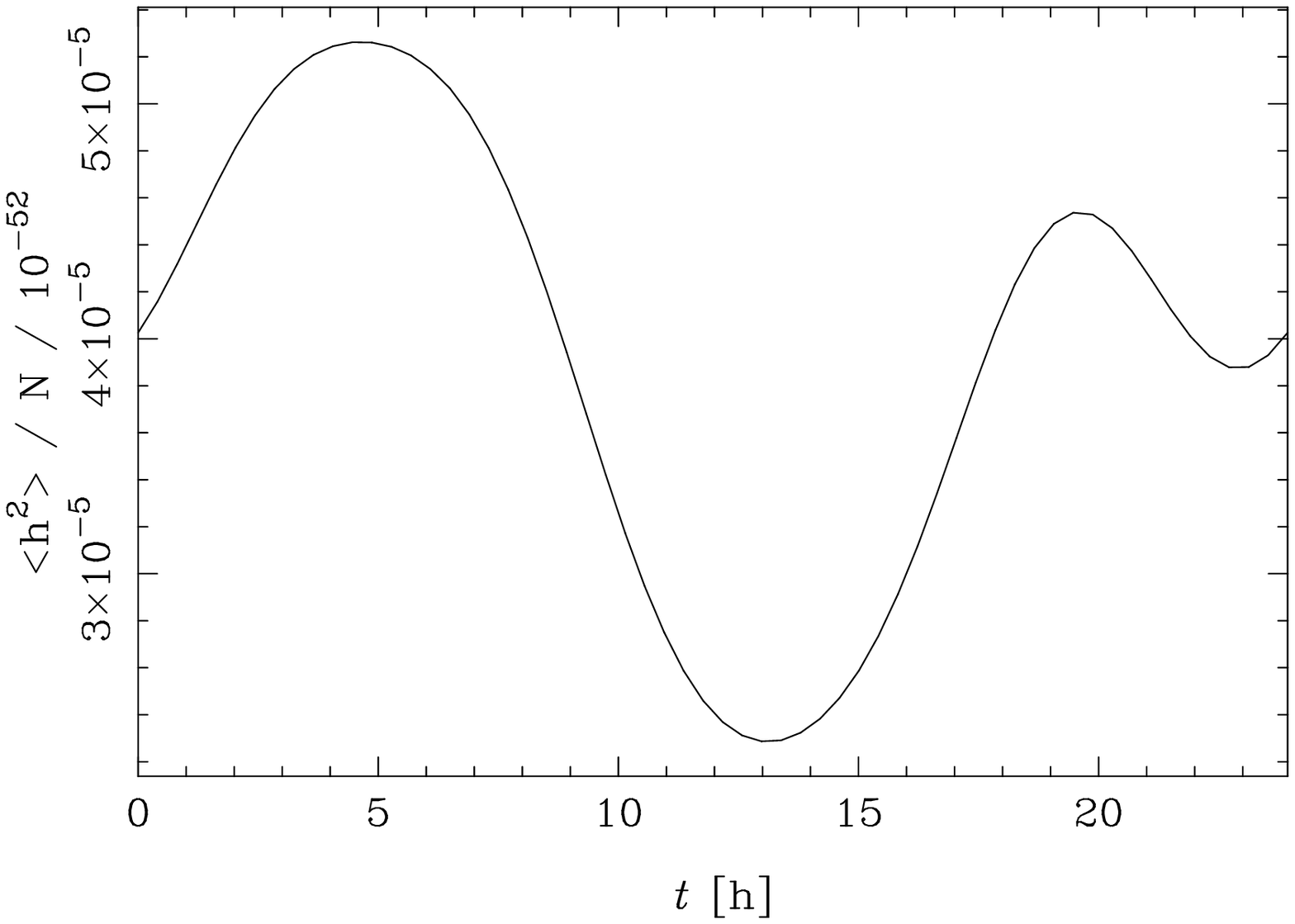,height=7.5cm} }
\caption[]{\label{f:disq}   \protect\footnotesize
Variation during one day of the 
total squared signal from a NS distribution concentrated in the Galactic disk,
according to Eqs.~(\ref{e:I(t):final}) and
(\ref{e:distrib:disq}) with the parameters 
$\overline{\epsilon^2} = (10^{-8})^2$, $\overline{I^2} = (10^{38}
{\rm\ kg\, m}^2)^2$, $\overline{P^{-4}} = (5{\rm\ ms})^{-4}$, 
$R_0 = 3.8{\rm\ kpc}$ and $z=0.5{\rm\ kpc}$.
The localization and orientation of the detector are those of VIRGO.
$t=0$ corresponds to a zero local sidereal time.}
\end{figure}

A Galactic disk distribution of NS can be modeled by the choice
\be \label{e:distrib:disq}
   f(r,\alpha,\delta) = \l\{ \begin{array}{lr}
 {\displaystyle	{\exp(-R/R_0) \ov 4\pi R_0^2 \, z_0 } }
	& \quad \mbox{if}\ |z| \leq z_0 \\
 	0 & \quad \mbox{if}\ |z| > z_0
   	\end{array} \r. \ , 
\ee
where $R=R(r,\alpha,\delta)$ is the distance to the Galactic rotation axis and
$z=z(r,\alpha,\delta)$ is the height above the galactic plane. 
Figure~\ref{f:disq} shows the value of $\langle  h^2(t) \rangle $ 
computed according to 
this distribution with $R_0 = 3.8{\rm\ kpc}$ and $z_0=0.5{\rm\ kpc}$. 

\begin{figure}
\centerline{ \epsfig{figure=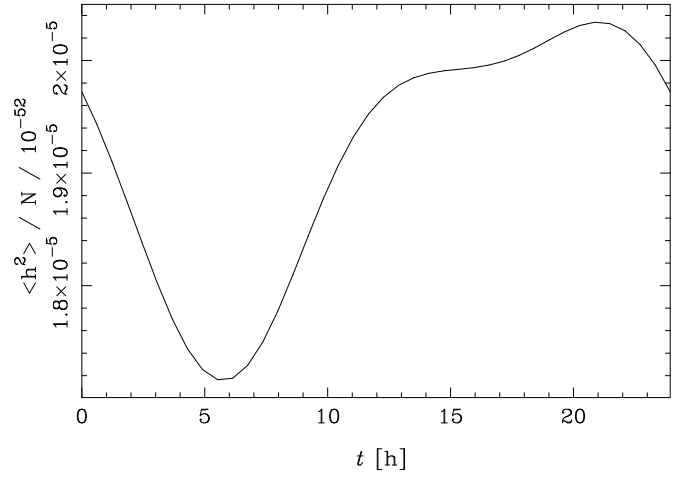,height=7.5cm} }
\caption[]{\label{f:halo}   \protect\footnotesize
Variation during one day of the 
total squared signal from a NS distribution corresponding to a Galactic halo,
according to Eqs.~(\ref{e:I(t):final}) and
(\ref{e:distrib:halo}) with the parameters 
$\overline{\epsilon^2} = (10^{-8})^2$, $\overline{I^2} = (10^{38}
{\rm\ kg\, m}^2)^2$, $\overline{P^{-4}} = (5{\rm\ ms})^{-4}$, 
and $a_0 = 10{\rm\ kpc}$.
The localization and orientation of the detector are those of VIRGO.
$t=0$ corresponds to a zero local sidereal time.}
\end{figure}

On the opposite, a 
NS distribution corresponding to a Galactic halo can be modeled by the choice
\be \label{e:distrib:halo}
 f(r,\alpha,\delta) = {64\ov 63\pi^2 a_0^3} \, { 
	(a / a_0)^2 \ov \sqrt{ (a_0/a) - 1} } \ , 
\ee
where $a = a(r,\alpha,\delta)$ is the distance from the Galactic centre
and $a_0$ is the radius of the halo.  
Figure~\ref{f:halo} shows the value of $\langle  h^2(t) \rangle $ 
computed according to 
this distribution with $a_0 = 10{\rm\ kpc}$.

\newpage

\newpage

\begin{table}
\caption[]{\label{t:hmax}
Gravitational radiation data for five selected pulsars. The GW amplitudes 
on Earth, $h_0$, are
computed according to  Eq.~(\ref{e:h0,num}) 
by assuming that $I=10^{38} {\ \rm kg\, m}^2$ 
(a representative value for a $1.4 \, M_\odot$ neutron star). 
$\epsilon_{\scriptscriptstyle -6}$ is the ellipticity in units of 
$10^{-6}$. 
The maximum ellipticity and maximum GW amplitudes 
are derived by attributing the totality of the observed pulsar spin-down 
rate to the emission of gravitation radiation.}
\begin{tabular}{cccccccrl}
 pulsar & distance & \multicolumn{2}{c}{GW}  & GW  &  
	maximum & maximum & \multicolumn{2}{c}{ellipticity}  \\
 name & & \multicolumn{2}{c}{frequencies} &amplitude & ellipticity 
	& GW amplitude & 
	\multicolumn{2}{c}{to get $h_0=10^{-26}$}  \\ 
  & $r {\ \rm [kpc]}$ & $f {\ \rm [Hz]}$ & $2f {\ \rm [Hz]}$ & $h_0$ & 
	$\epsilon_{\rm max}$ & $h_{0,\rm max}$ & 
	\multicolumn{2}{c}{$\epsilon_{\rm detect}$} \\ 
\tableline
Vela & 0.5 & 11 & 22 & $1.1\times 10^{-27}\, \epsilon_{\scriptscriptstyle -6}$ &
	$1.8\times 10^{-3}$ & $1.9\times 10^{-24}$  
	& $9.1\times 10^{-6}$  & $=5\times 10^{-3} \, \epsilon_{\rm max}$ \\
Crab & 2 & 30 & 60 & $1.9\times 10^{-27} \, \epsilon_{\scriptscriptstyle -6}$ &
	$7.5\times 10^{-4}$ & $1.4\times 10^{-24}$  
	& $5.3\times 10^{-6}$  & $=7\times 10^{-3} \, \epsilon_{\rm max}$ \\
Geminga & 0.16\tablenote{this is the recently determined distance from the
measure of Geminga parallax \cite{CaBMT96}}
 & 4.2 & 8.4 & $4.7\times 10^{-28}\, 
	\epsilon_{\scriptscriptstyle -6}$ & 
	$2.3\times 10^{-3}$ & $1.1\times 10^{-24}$ 
	& $2.1\times 10^{-5}$ & $=9\times 10^{-3} \, \epsilon_{\rm max}$ \\ 
\tableline
B1957+20 & 1.5 & 621 & 1242 & $1.1\times 10^{-24} \, 
	\epsilon_{\scriptscriptstyle -6}$ &
	$1.6\times 10^{-9}$ & $1.7\times 10^{-27}$  
	& $9.1\times 10^{-9}$  & $> \epsilon_{\rm max}$ \\
J0437-4715\tablenote{ref. \cite{Johns93}} & 0.14 & 174 & 348 & 
	$9.1\times 10^{-25}\, 
	\epsilon_{\scriptscriptstyle -6}$ &
	$2.9\times 10^{-8}$ & $2.6\times 10^{-26}$  
	& $1.1\times 10^{-8}$  & $=0.4 \, \epsilon_{\rm max}$ \\
\end{tabular}
\end{table}

\begin{table} 
\caption{\label{tab1} 
Fourier spectrum of the solar flux at the
latitude of Pisa.}
\begin{tabular}{cccc}
 frequency & period & identification &  spectral component \\
  $\nu$ [Hz] & $\nu^{-1}$ [h] & & (arbitary units) \\
\tableline
   3.1688089E-08 &   8766.000   & 1 year &    0.6391195  \\  
  6.3376177E-08 &  4383.000     & 6 months & 1.4378662E-02	\\
  1.1510698E-05 &  24.13214     & $\nu_{\rm sol} - 2 \nu_{\rm ty}$
							& 4.0333651E-02	\\
  1.1542386E-05 &  24.06589  &  $\nu_{\rm sol} - \nu_{\rm ty}$  &  0.4005030  \\
  1.1574074E-05 &  24.00000  & $\nu_{\rm sol}$ (solar day) &    1.836788    \\
  1.1605762E-05 & 23.93447   & $\nu_{\rm sol} + \nu_{\rm ty}$ (sidereal day)
					 	& 0.4006524    \\
  1.1637450E-05 &  23.86930  & $\nu_{\rm sol} + 2 \nu_{\rm ty}$ 
							& 3.9582606E-02	\\
  2.3084773E-05 &  12.03294  &  $2\nu_{\rm sol} - 2\nu_{\rm ty}$   	
						& 5.2575748E-02	\\
  2.3116459E-05 & 12.01645   & $2\nu_{\rm sol} - \nu_{\rm ty}$  
						& 1.2078404E-02	\\
  2.3148148E-05 &  12.00000  &  $2\nu_{\rm sol}$ 	& 0.7087801    	\\
  2.3179837E-05 &  11.98359  &  $2\nu_{\rm sol} + \nu_{\rm ty}$ 	
						& 1.4178075E-02	\\
  2.3211524E-05 & 11.96724   &   $2\nu_{\rm sol} + 2 \nu_{\rm ty}$	
							& 5.2661020E-02	\\
\end{tabular}
\end{table}

\begin{table} 
\caption{\label{t:fourier:h2}
Fourier spectrum of the total squared signal $\langle  h^2(t) \rangle $
corresponding to a disk distribution of NS and depicted in 
Fig.~\ref{f:disq}.}
\begin{tabular}{ccc}
 frequency & cosine coef. & sine coef.   \\
   & (arbitary units) & (arbitary units) \\
\tableline
 0	& 7.895	& 0 \\
 $\nu_{\rm sid}$ & 0.923 & 0.593 \\
 $2 \nu_{\rm sid}$ & -0.738  & -0.065 \\
 $3 \nu_{\rm sid}$ & -0.079 & 0.124 \\
 $4 \nu_{\rm sid}$ & 0.005  & 0.087  
\end{tabular}
\end{table}

\end{document}